\documentclass[sigconf]{acmart}

\usepackage{csquotes}
\usepackage{hyperref}
\usepackage{subcaption}

\usepackage{xcolor}
\usepackage{soul}

\newcommand{\labelphantom}[1]{%
  \parbox{0pt}{\phantomsubcaption\label{#1}}%
}
\newcommand{\eg}{\textit{e.g.}}
\newcommand{\ie}{\textit{i.e.}}

\newcommand{\etal}{\textit{et al.}}

\newcommand{\leftcell}[2][l]{%
  \begin{tabular}[#1]{@{}l@{}}#2\end{tabular}}

\AtBeginDocument{%
  \providecommand\BibTeX{{%
    \normalfont B\kern-0.5em{\scshape i\kern-0.25em b}\kern-0.8em\TeX}}}

\copyrightyear{2024}
\acmYear{2024}
\setcopyright{rightsretained}
\acmConference[CHI '24]{Proceedings of the CHI Conference on Human Factors in Computing Systems}{May 11--16, 2024}{Honolulu, HI, USA}
\acmBooktitle{Proceedings of the CHI Conference on Human Factors in Computing Systems (CHI '24), May 11--16, 2024, Honolulu, HI, USA}
\acmDOI{10.1145/3613904.3642179}
\acmISBN{979-8-4007-0330-0/24/05}

\begin{document}

\title[AI-Assisted CPD for HCD]{AI-Assisted Causal Pathway Diagram for Human-Centered Design}

\author{Ruican Zhong}
\authornote{Both authors contributed equally to this research.}
\orcid{0009-0004-7169-0675}
\email{rzhong98@uw.edu}
\affiliation{%
  \institution{University of Washington}
  \city{Seattle}
  \state{WA}
  \country{USA}}
\author{Donghoon Shin}
\authornotemark[1]
\orcid{0000-0001-9689-7841}
\email{dhoon@uw.edu}
\affiliation{%
  \institution{University of Washington}
  \city{Seattle}
  \state{WA}
  \country{USA}}
\author{Rosemary Meza}
\orcid{0000-0002-4849-7250}
\email{rosemary.x1.meza@kp.org}
\affiliation{%
  \institution{Kaiser Permanente Washington Health Research Institute}
  \city{Seattle}
  \state{WA}
  \country{USA}}
\author{Predrag Klasnja}
\orcid{0000-0002-4570-703X}
\email{klasnja@umich.edu}
\affiliation{%
  \institution{University of Michigan}
  \city{Ann Arbor}
  \state{MI}
  \country{USA}}
\author{Lucas Colusso}
\orcid{0000-0002-9773-3483}
\email{lucascolusso@microsoft.com}
\affiliation{%
  \institution{Microsoft}
  \city{Redmond}
  \state{WA}
  \country{USA}}
\author{Gary Hsieh}
\orcid{0000-0002-9460-2568}
\email{garyhs@uw.edu}
\affiliation{%
  \institution{University of Washington}
  \city{Seattle}
  \state{WA}
  \country{USA}}

\renewcommand{\shortauthors}{Zhong \& Shin et al.}

\begin{abstract}
This paper explores the integration of causal pathway diagrams (CPD) into human-centered design (HCD), investigating how these diagrams can enhance the early stages of the design process. A dedicated CPD plugin for the online collaborative whiteboard platform Miro was developed to streamline diagram creation and offer real-time AI-driven guidance. Through a user study with designers ($N=20$), we found that CPD's branching and its emphasis on causal connections supported both divergent and convergent processes during design. CPD can also facilitate communication among stakeholders. Additionally, we found our plugin significantly reduces designers' cognitive workload and increases their creativity during brainstorming, highlighting the implications of AI-assisted tools in supporting creative work and evidence-based designs.
\end{abstract}

\ccsdesc[500]{Human-centered computing~Systems and tools for interaction design}
\ccsdesc[500]{Human-centered computing~HCI design and evaluation methods}

\keywords{causal pathway diagram, human-centered design, generative AI, LLM, implementation science}

\maketitle

\section{Introduction}

Causal pathway diagram (CPD) is a graphical tool that represents the causal relationships between variables and the desired outcomes under a specific context~\cite{lewis2018classification, cpdtoolkit}. CPD is a valuable tool in designing theory-driven behavioral implementation strategies~\cite{lewis2018classification}. CPD uses graphical representations to depict the mechanisms in which an implementation strategy is thought to work~\cite{cpdtoolkit}. It helps map out how an implementation strategy can achieve desired outcomes; highlighting factors that are necessary or helpful to achieve those outcomes.

\begin{figure*}[h!]
    \centering
    \includegraphics[width=.85\textwidth]{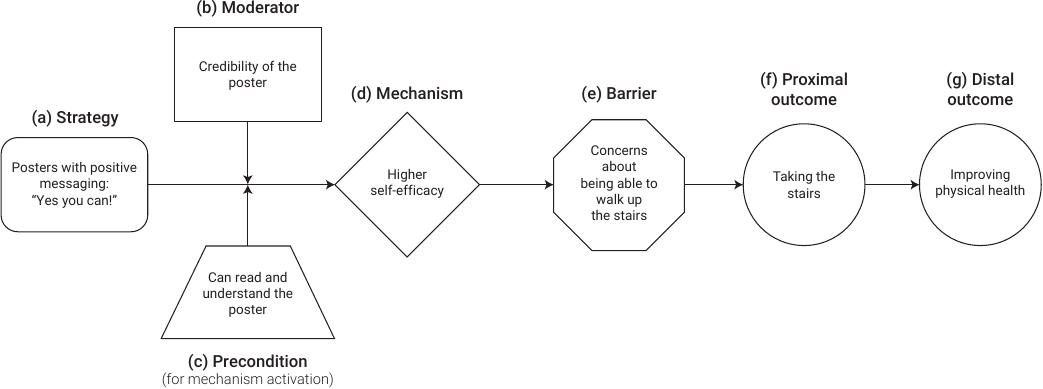}
    \caption{An example of causal pathway diagram of increasing clients' physical activities}
    \label{fig:cpd_example}
    \labelphantom{fig:cpd_example:strategy}
    \labelphantom{fig:cpd_example:moderator}
    \labelphantom{fig:cpd_example:precondition}
    \labelphantom{fig:cpd_example:mechanism}
    \labelphantom{fig:cpd_example:barrier}
    \labelphantom{fig:cpd_example:proximal_outcome}
    \labelphantom{fig:cpd_example:distal_outcome}
\end{figure*}

Consider the scenario where designers are tasked with increasing clients' physical activities (\autoref{fig:cpd_example}). They can use CPD to synthesize their own user research with published research to map out the factors that can lead to this desired outcome (\autoref{fig:cpd_example:distal_outcome}). Working backward, they can map out relevant proximal goals (\eg{}, having clients take the stairs instead of the elevators; \autoref{fig:cpd_example:proximal_outcome}), identify barriers that prevent target clients from taking the stairs (\eg{}, concerns about not being able to walk up all the stairs; \autoref{fig:cpd_example:barrier}) and appropriate mechanisms (\eg{}, increase self-efficacy; \autoref{fig:cpd_example:mechanism}) to help overcome the barriers. They may also use CPD to consider relevant moderators and preconditions they must consider during design. For example, if the design strategy they choose to focus on is a low-tech solution of displaying a poster with positive messaging of \textit{``Yes you can!''} to help increase clients' belief that they can take the stairs, designers may note moderators (\eg{}, the credibility of the poster; \autoref{fig:cpd_example:moderator}) and preconditions (\eg{}, whether users can read and understand the poster; \autoref{fig:cpd_example:precondition}) that is important for the design to function as intended. CPD can also support brainstorming. Designers can consider other paths that can result in the same distal outcomes, potentially exploring alternative proximal outcomes, barriers, mechanisms, or strategies. Finally, with this visualization, designers can articulate the design goals and why the overall strategy may work for various stakeholders (\autoref{fig:cpd_example:strategy}). The use of CPDs can be valuable throughout a design project: at the start of the project by identifying relevant factors and mechanisms to prioritize and optimize; during the implementation by addressing barriers; and at the end of the project to help diagnose and evaluate outcomes~\cite{lewis2022mechanics}.

Though originally developed for implementation science, CPD may be extended to human-centered design (HCD), as HCD and implementation science share similar underlying objectives. At the root of both implementation science and human-centered design is the question of how to ensure that \textit{what} we design will ultimately \textit{work well}; be used and adopted by users to achieve the desired outcome. There has been increasing interest both within the HCI community and in the implementation science community to strengthen the exchange of methods and integration of ideas \cite{lyon2020leveraging, dopp2020aligning, lyon2023bridging}. In the same way that an implementation scientist may use CPD to help design and evaluate implementation strategies, designers may also find value in using CPD in the earlier stages of design as a way to drive evidence-based thinking. Designers may use CPD to model their synthesized research findings into a visual representation of users' goals and barriers, and identify relevant contextual factors that can influence design uptake. They may also use CPD to guide ideation and the development of design hypotheses.    



However, despite the potential benefits of CPD, several challenges may exist that can prevent designers from using and integrating CPD into their work. Conceptually, the use of CPD involves thinking about the interventions and strategies that are being designed at a different level of abstraction and granularity than may come naturally to design practitioners~\cite{lewis2018classification}. Instead of directly diving into brainstorming, when using CPD, designers are challenged to first identify and organize relevant factors in an evidence-based way. This different design paradigm may be perceived as difficult to integrate into existing workflows. Further, those who are unfamiliar with CPD may find it hard to adapt to the CPD components and syntax (\ie{}, what is a ~\textit{mechanism} and how should it be represented). Finally, at a practical level, generating these causal diagrams can incur additional costs in drawing, sharing, and iterating with peer feedback.

In order to address these potential challenges of adopting CPD in HCD, we developed a CPD plugin for an online collaborative whiteboard, Miro. Specifically, the plugin provided CPD elements that can be used directly by the designers. This enables people to focus on the conceptual work in CPD and minimize time spent on making the visual diagrams. The plugin also included features that provide guidance on how to develop CPDs and provide real-time suggestions using generative AI (\ie{}, GPT-4) to support an exploration of relevant factors. 

We conducted a user study with 20 design practitioners. Participants were asked to brainstorm solutions for design sprints with the help of CPD and the plugin. Specifically, we addressed the following research questions:
\begin{itemize}
   \item \textbf{RQ1}: How do designers use CPD in human-centered design?
   \item \textbf{RQ2}: How does the plug-in support practitioners in designing with CPD?
\end{itemize}

We found that designers are positive about integrating CPD in the early stages of human-centered design, especially for brainstorming and strategic prioritization. Designers pointed out that CPD is particularly effective because it emphasizes directly addressing the root cause of the problem as a goal-oriented design process. We also found that our plugin was able to reduce the cognitive workload involved in memorizing the CPD framework itself, allowing people to focus their energy on the design task. Furthermore, with LLM-generated recommendations, our plugin was able to enhance designers' creativity in brainstorming sessions. However, designers also emphasized the potential consequences of using the AI suggestions irresponsibly and highlighted opportunities to make the AI recommendations more evidence-based and to communicate the information provenance.

In this work, we offer several important contributions:

\begin{itemize}
    \item Introduction and study of how practitioners can use CPD in the early stages of human-centered design
    \item A diagramming plugin\footnote{The codebase and the application of our plugin are available on the following links: (i) \href{https://github.com/prosociallab/chi2024\_miro-cpd-plugin}{codebase}, (ii) \href{https://miro.com/app-install/?response\_type=code\&client\_id=3458764542648826179\&redirect\_uri=\%2Fconfirm-app-install\%2F}{plugin (application)}} that helps designers generate and iterate on CPDs to guide design
    \item Insights about the challenges and opportunities of using AI assistance in supporting creative work and evidence-based designs
\end{itemize}

\section{Related Work}
\subsection{Theory Use in Human-Centered Design}
\label{related-work:HCD}
Human-centered design (HCD) is built on the idea that through a better understanding of people, we can generate more effective designs~\cite{shneiderman2010designing,rogers2002interaction,carroll2003hci,benyon2013designing}. There has been a long tradition of using theory to guide design~\cite{rogers2004new}, such as the use of cognitive science theories to guide the design of computer interfaces~\cite{carroll1991designing}, use of social psychology theories to design online communities~\cite{kraut2012building}, and the use of social and behavior theories to guide behavior change technologies~\cite{consolvo2009theory}. 

Through their synthesis of prior research~(\eg{},~\cite{rogers2004new, rogers1hci, bederson2003craft}), van Turnhout et al. points out six key types of functions of theory when designing technologies for people~\cite{van2019practical}:
(1) Description, to identify phenomena and describe them in a consistent and clear manner; (2) Explanation, to expose underlying causes and their relationships; (3) Generative, to support the generation of novel ideas and design alternatives; (4) Aspirational, to explicate what ideals, goals or values to strive for in design; (5) 
Prediction, to predict effects in normal and novel situations; and (6)
Prescription, to provide advice about what (not) design such as guidelines, and best practices.

Within a typical human-centered design process, these functions may manifest in different ways. For example, during brainstorming, theories can help by suggesting ideas for designers to explore. They can support divergent thinking by identifying space to explore. They could make salient how relevant strategies influence users and behaviors, and highlight barriers to overcome. They can help designers generate design hypotheses about the outcomes of their designs. Theories also help in convergent stages of design by guiding decisions on which designs to prototype and implement so that the design goals can be optimized. During evaluation, theories can inform study design and help interpret findings. At a more meta level, theories can support communication with stakeholders, both to explain the goal of design as well as to help justify and get buy-in for design decisions.  

However, despite the various benefits of using theory in design practice~\cite{hekler2013mind}, many have noted that research and practice gaps exist~\cite{colusso2019translational, norman2010research, gray2014reprioritizing, roedl2013design, rogers2004new}. Effectively using theories to approach design problems has faced unique and enduring challenges~\cite{dourish2006implications, schon1983reflective}. There are several critical barriers preventing their usage. First, designers are often not trained in relevant basic science fields,  and consuming and expressing knowledge in a theory-driven, or evidence-based approach can be difficult~\cite{beck2016examining, colusso2017translational}. Second, even if designers are familiar with relevant theoretical insights, how to translate theoretical insights into the ``messy'' real-world context may not be simple and straightforward~\cite{rogers2004new}. Finally, when to engage theory may also be unclear. How to incorporate theories in a format that can fit with existing design processes is also important for theory usage~\cite{colusso2017translational,colusso2018behavior}.

\subsection{Causal Pathway Diagramming (CPD)}

In this paper, we hypothesize that causal pathway diagramming (CPD) may be a useful tool to guide theory-driven human-centered design. CPDs are box-and-arrow diagrams that depict ``interrelations among variables and outcomes of interest in a given context.''~\cite{lewis2018classification} CPDs have been proposed in implementation science as a way to support evidence- or theory-based design, to help overcome the observation that most existing health care and public health implementation fail to achieve the intended change~\cite{damschroder2009fostering}.  By using CPDs to map out the ways in which interventions could overcome barriers and how they could improve the outcomes of interest, CPDs could be applied across different stages of HCD, to help: (1) inform the design and development of the strategy, (2) support the brainstorming of new strategies, (3) increase the impact of existing strategies, and (4) help prioritize which strategies to use in which contexts~\cite{lewis2018classification}.

Though CPD has only been recently proposed in health implementation, there has been a long-established history of using these types of diagrams across various fields. At its abstract form, causal pathways graphically describe causal relationships within a set of variables, and are widely used in social and behavioral sciences and statistics~\cite{pedhazur1982multiple}. More recently, the Theory of Change (ToC)~\cite{taplin2013theory} has also used the Outcomes Framework as a way to provide a visual representation of the preconditions and requirements necessary to achieve a desired goal. The idea is that by defining the outcome of interest and identifying rationales and assumptions, the Outcomes Framework can help philanthropies and nonprofits plan and evaluate systems in a more evidence-based way~\cite{connell1998applying}. However, while many have proposed this type of causal mapping to support human-centered design, its use is still relatively limited. Outside of global health where there has been an intersection of methods from ToC and HCD~\cite{lafond2021theory}, this type of causal pathway diagramming is rarely referenced as part of the design process in human-centered design and HCI literature. Thus, one of the research questions is to explore the potential benefit of using this type of visual representation in human-centered design. 

In our work, we use the CPD method that has been developed by the OPTICC Center~\cite{lewis2021optimizing}, a National Cancer Institute (NCI) funded center focused on optimized evidence-based intervention implementation. Unlike the Outcomes Framework from (ToC)~\cite{taplin2013theory}, OPTICC's CPD method puts forward a formal syntax~\cite{cpdtoolkit}. First is the \textit{strategy}, or intervention of focus. It is depicted as a rounded rectangle, and is typically the leftmost element of a CPD. What follows a strategy element is usually a \textit{mechanism} element, which explains how or why the strategy works. Mechanisms are depicted as diamonds. Another key component of the CPD is the \textit{barrier}, or the obstacle in place that prevents the achievement of the desired outcome. Barriers are depicted as octagons. Outcomes are depicted as circles and are typically the last element in a CPD. There could be multiple versions of the outcome circles, including \textit{proximal outcomes}, \textit{intermediate outcomes}, and \textit{distal outcomes}. Outside of the stem of the CPD, there are also moderators and preconditions. Moderators are depicted by rectangles, and represent factors that facilitate or impede a part of the causal process. Preconditions are depicted as isosceles trapezoids, and represent factors that are necessary for a part of the causal process. 
While having more formalized syntax can help structure and improve the consistency of causal diagrams, it can pose a barrier to using CPDs. This is on top of needing to learn what causal pathways are and how to use them. Therefore, our second research question is whether we can build a tool to facilitate the use of causal pathways in design.

\subsection{Potential Use of CPD During Ideation}

Many tools have been designed to support creativity in HCD ~\cite{frich2019mapping, wang2017literature, gabriel2016creativity}. However, few of these creativity support tools (CSTs) have been developed with an emphasis on supporting the generation of ideas in an evidence or theory-driven way. One set of CSTs simply stimulates designers to think creatively. These tools are not specific to the design context and often just contain inspirational images or words\mbox{~\cite{eameshouseofcards1952,hsieh2023cards}}, or provocative concepts ~\cite{vines2012questionable}. Others support brainstorming by suggesting additional relevant ideas based on a set of existing ideas inputted by the user\mbox{~
\cite{clark2018creative, wan2023investigating, bunian2021vins, feng2022gallery, lu2024ai, wang2010idea, andolina2015inspirationwall, andolina2017crowdboard, ferdowsi2023brainstorming}}. Often, these tools help surface inspirational stimuli by varying the analogical distance, commonness, and modality of the example ideas \mbox{~
\cite{chan2011benefits, lawton2023drawing, zhang2022storydrawer, cai2023designaid, zamfirescu2023towards}}. IdeaExpander, for example, draws on the conversations of group brainstorming and provides recommendations of related pictures to stimulate a divergent brainstorming process\mbox{~\cite{wang2010idea}}. Clark et al.'s work on creative writing suggests related information based on existing user inputs, which helped with ideating possible slogans and stories\mbox{~\cite{clark2018creative}}. VINS is a system that recommends relevant UI examples to support UI design brainstorming\mbox{~\cite{bunian2021vins}}. While these tools could help designers come up with more innovative ideas more efficiently, without the integration of evidence from user research or published research, it is unclear if the generated ideas will work when addressing the design problem.

Integrating CPD into creativity support may help scaffold the designers to consider the relevant evidence or theory throughout--and thus ensuring a higher likelihood of the generated ideas to be effective\mbox{~\cite{lewis2022mechanics}}. CPD can also be useful during the four key types of CST activities as noted by Shneiderman: collect, relate, create, and donate\mbox{~\cite{shneiderman2002creativity}}. Collecting refers to the process of gathering insights from prior literature and resources. The creation of a CPD would encourage designers to collect and utilize research findings and theories in order to examine relevant causal pathways. Relating refers to consulting with their colleagues and managers throughout the design process. CPD, as a graphical representation of how a possible design solution addresses the problem space, could serve as the microtheory to facilitate the discussions of goals and relevant factors to consider. Creating refers to the divergent thinking process of exploring multiple solutions. CPD could allow designers to explore different, evident-based paths, providing a way to explore possible solutions. Donating means distributing the results publicly. The generated CPDs could provide a structure for designers to store and share the design solutions they have explored, and refer to the underlying factors (\eg{}, mediators, moderators, and mechanisms) that have contributed to those ideas.

\subsubsection{AI-assisted creativity support tools}

Thanks to the ability of AI to model data and extract knowledge~\cite{hwang2022too, kun2019creative}, creativity support tools have explored the use of AI in their design. One line of research is on using machine learning models fine-tuned to a \textit{specific domain} to support a specific type of design and ideation task (\eg{},~\cite{wan2023investigating, bunian2021vins, clark2018creative, feng2022gallery, lu2024ai, wang2010idea}). For example, VINS uses an attention-based neural network to retrieve related UI examples from a given collection to support the brainstorming process\mbox{~\cite{bunian2021vins}}, and recent work by Wan~\etal{}\mbox{~\cite{wan2023investigating}} used GAN (generative adversarial network) trained with relevant research data to support visual brainstorming and help designers explore the semantic space of diagramming ideation. Despite the efficacy of such approaches, however, they are limited in that the set of evidence or resources these AI-assisted CSTs could draw on is quite limited. As most machine learning systems like these CSTs are trained specifically for one domain, they cannot easily generalize to other contexts. In addition, the evidence they use is often restricted in number or the range of content, as the sets were either hand-curated or drawn from a specific online resource.

Recently, the advance of large language models (LLMs) has opened up new opportunities to address this space~\cite{joosten2024comparing, lawton2023drawing, jeon2021fashionq, ko2023large, gero2019stylistic, mirowski2023co, yuan2022wordcraft, kim2023diarymate, wan2023investigating, petridis2023anglekindling, york2023evaluating}. Compared to existing, task-specific AI techniques, LLMs are not constrained by domain or by the set of resources they can use to generate suggestions. Trained on a large corpus of data (\eg{}, newspapers, academic papers, blogs, etc.) with human feedback, LLMs enable users to leverage their large knowledge base to make suggestions in many domains\mbox{~\cite{bommasani2021opportunities}}. Among various application areas, one promising application of LLMs is to support users in iterating and improving on existing ideas, allowing them to explain their ideas in more detail~\cite{gero2019metaphoria, lawton2023drawing, kim2023metaphorian, hou2024c2ideas, zamfirescu2023towards, epstein2022happy, almeda2023prompting,druga2023scratch, di2022idea, yuan2022wordcraft}. For instance, C2Ideas uses LLMs to generate iterations of the descriptions of users' design intentions, highlighting possible design directions by clarifying users' inputs~\cite{hou2024c2ideas}. In our work, we posit that LLMs could possibly support the creation of CPD by making suggestions that help designers better articulate their ideas. Another set of work discussed how generative AI could be used to provide a wide range of suggestions based on the users' existing ideas, augmenting creativity~\cite{ko2023large, petridis2023anglekindling, wan2023felt, rick2023supermind, druga2023scratch}. These works also highlight the potential of LLMs in providing recommendations that motivate designers to think divergently, increasing the number of design directions generated~\cite{jeon2021fashionq, angert2023spellburst, mccaffrey2018human, klein2020beyond}. For instance, AngleKindling uses LLMs to suggest different angles to interpret a press release, helping journalists iterate on their writings~\cite{petridis2023anglekindling}. This demonstrates the potential of LLMs in supporting creativity, especially by bringing in additional perspectives or related literature that might be relevant to the design space, encouraging evidence-based design thinking.

Overall, recent works on LLMs and their applications demonstrate the promising potential to support creativity. Our work builds on these lines of research to explore the use of LLMs in supporting the creation of CPDs.
\section{Design of the Plugin}\label{section:design}

To streamline the process of creating and iterating on CPDs and support the designers throughout the process, we developed a plugin in Miro~\cite{miro}, a widely used diagramming platform. Miro lets users create diverse visual elements, such as text boxes, circles, and rectangles on a collaborative board. It is also often used in the early stages of HCD for brainstorming and ideation purposes. The simplicity of interactions and design elements within Miro, coupled with its plugin-friendly architecture, made it an appropriate platform for deploying and testing our concept--how to support designers in generating CPDs.

Below we describe the design and implementation details of the key features embedded in our plugin. Our plugin uses a multi-tab design, where each feature of the plugin can be used separately and independently. We provide a more detailed visual walkthrough of each component in \autoref{appendix:keyscreens}.

\subsection{Key Features of the Plugin}

\begin{figure*}[ht]
    \centering
    \includegraphics[width=\linewidth]{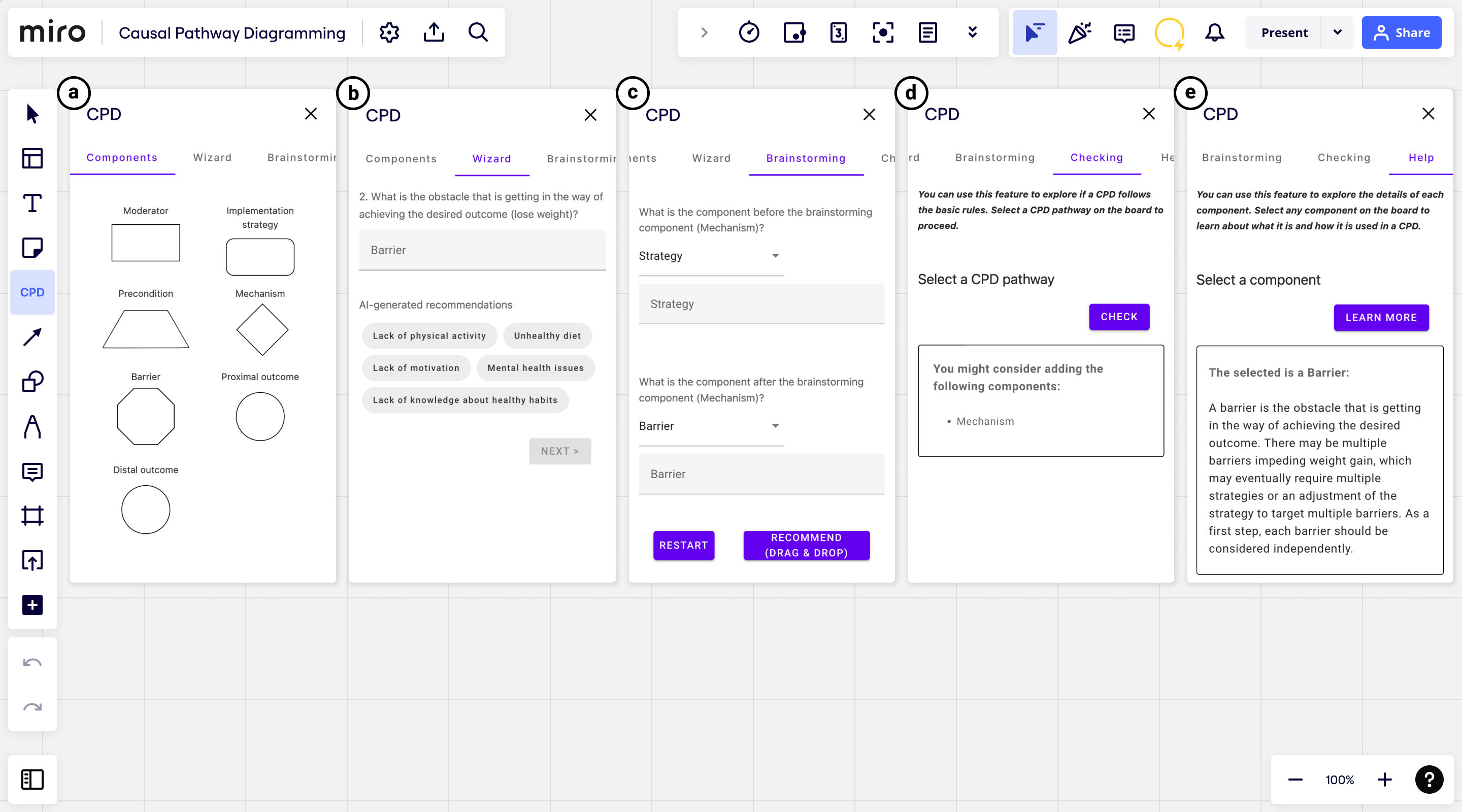}
    \caption{Key screens of our plugin. Consisting of five features, the plugin streamlines the creation and validation of CPD}
    \label{fig:keyscreen}
    \labelphantom{fig:keyscreen:component}
    \labelphantom{fig:keyscreen:wizard}
    \labelphantom{fig:keyscreen:brainstorming}
    \labelphantom{fig:keyscreen:checking}
    \labelphantom{fig:keyscreen:help}
\end{figure*}

\subsubsection{Component}

Using this feature, users could easily create the elements of CPDs on the board: moderator (rectangle), implementation strategy (round rectangle), precondition (trapezoid), mechanism (rhombus), barrier (octagon), proximal outcome (circle), and distal outcome (circle). \autoref{fig:keyscreen:component} displays the layout of this feature. 

For each item shown in the panel, users can easily drag and drop them onto the Miro board. Once the user drags and drops an element, the plugin creates an empty entity complemented by a hyperlinked title that directs users to the page that contains the definition of the entity. This not only facilitates quick access to the definition but also streamlines the process of element creation on the Miro platform, sparing users the hassle of manually having to locate the shape in the platform. Plus, the plugin shows the definition of the element in a tooltip when the user hovers their mouse over the element.

\subsubsection{Wizard}

To streamline the creation of CPD, we designed and developed a form-based wizarding tool that guides the user in a step-by-step manner (\autoref{fig:keyscreen:wizard}). This functionality navigates users through the CPD generation process by eliciting inputs for each component. We guide the creation using backward mapping~\cite{elmore1979backward}, where participants focus on the distal outcome first, then work backward to the barrier, proximal outcome, strategy, and mechanism. For each component, the user is provided with an explanation of the component (\eg{}, Barrier - \textit{"What is the obstacle that is getting in the way of achieving the desired outcome?"}).

One of the challenges with the creation of CPD may be a lack of knowledge about relevant theory. To overcome this challenge, we sought to provide relevant component recommendations to users. In the long run, our plan is to establish a CPD repository and to recommend factors from our repository. However, until such a repository is well populated, we sought an innovative use of LLM to provide recommendations. We hypothesized that LLMs, building on a large body of data, may be able to provide a set of contextually relevant factors given other already inputted components~\cite{wu2023survey, hou2023large}.

Thus, we designed the \textit{Wizard} such that starting from the desired distal outcome, after entering the content of each component, the system recommends to the user up to five candidates for each of the following components by leveraging an LLM. To implement this AI-generated recommendation, we prompted the LLM with careful instructions. The following is an example prompt for generating recommendations for a proximal outcome based on users' input of distal outcome. The structure of the prompts is similar throughout the step-by-step process.

\begin{displayquote}
\texttt{Based on the \{previous element(s)\} the user have input, recommend 5 possible \{current element\}:\\\hspace*{1mm}- \{the first previous element\}: \{content for the first previous element\}\\\hspace*{1mm}- \{the second previous element\}: \{content for the second previous element\}\\\hspace*{1mm}...}
\end{displayquote}

Once the user finalizes all components, users can easily drag \& drop to a specific area on the board, which creates a complete CPD populated with their inputs on a designated position.

\subsubsection{Brainstorming}

Our plugin also provides users with the ability to explore potential candidates for a specific component (\autoref{fig:keyscreen:brainstorming}) using LLM. This is designed to help support the branching from a specific component in CPDs. For instance, knowing what the barrier is in a CPD, users may be interested in exploring different possible mechanisms to address that barrier, expanding their innovative thinking.

To use this feature, users need to select the specific component they intend to brainstorm first. Then the plugin proceeds to request information regarding the contents of the preceding and/or following components related to the chosen element. With these contextual inputs, the plugin generates and recommends candidates for the component.

Similar to the approach we took for implementing wizarding, we instructed an LLM with the following prompt:

\begin{displayquote}
\texttt{Based on the \{preceding element\} and \{following element\} the user has input, recommend 5 possible \{element that the user wishes to brainstorm\}:\\\hspace*{1mm}- \{preceding element\}: \{content for the preceding element\}\\\hspace*{1mm}- \{following element\}: \{content for the following element\}}
\end{displayquote}

\subsubsection{Checking}

In contrast to the previous features that primarily aim to initially create or brainstorm CPD components, the checking feature is designed to help users verify the CPDs they have already generated (\autoref{fig:keyscreen:checking}). Specifically, our checking feature lets users \textit{diagnose} if any of the following issues are present in their CPD:

\begin{itemize}
    \item One or more required elements are missing
    \item One or more elements are not connected to the CPD pathway
    \item One or more elements are connected in the wrong order
    \item The CPD does not start / end with implementation strategy / distal outcome
\end{itemize}

This feature mainly focuses on checking the basic correctness of a CPD. To start, users should select a generated CPD on the board, including its components and connections between components. Then, users should click on the ``check'' button to perform checking. If the selected CPD does not contain any of the listed issues, we provide positive confirmation such as \textit{``No syntax issues with your CPD pathway!''} If the selected CPD contains one or more of these issues, we provide suggestive feedback about the specific problems, such as ``\textit{You might consider adding the following components.}''

\subsubsection{Help / Glossary}

Finally, the help feature allows users to easily access the definition of an element that is already on the board (\autoref{fig:keyscreen:help}). By selecting an existing element on the board and clicking on the ``learn more'' button, the plugin promptly displays the element's definition (\ie{}, strategy - \textit{``Strategy is an element that the diagram is intended to unpack. It is important to make the strategy concrete, to write it as it would be performed in that particular setting.''})

\subsection{Implementation}

Our system was built on a Javascript-based framework (SvelteKit), and deployed on the Miro SDK 2.0, which allowed the system to interact with the Miro board (\eg{}, add/detect items on the board).

To generate every LLM-generated output, we used GPT-4 with the following parameters: \texttt{temperature}: 1, \texttt{max\_tokens}: 256, \texttt{top\_p}: 1, \texttt{frequency\_penalty}: 0, and \texttt{presence\_penalty}: 0. To avoid potential hallucination issues which are frequently experienced by LLMs, we provided the model with the definition of the elements of CPDs, by prepending the definitions to the prompt, whenever the model is invoked. We used GPT-4 without further tuning because existing research showed that a generic model is already able to achieve good performance in domain-specific tasks~\cite{bubeck2023sparks} and that GPT-4 can provide domain-specific suggestions with strategized prompting~\cite{Nori2023CanGF}.
\section{User Study}

To understand the use of CPD and our plugin to support human-centered design, we conducted a user study with design practitioners.

\subsection{Procedure}

\begin{figure*}[t]
    \centering
    \includegraphics[width=.9\linewidth]{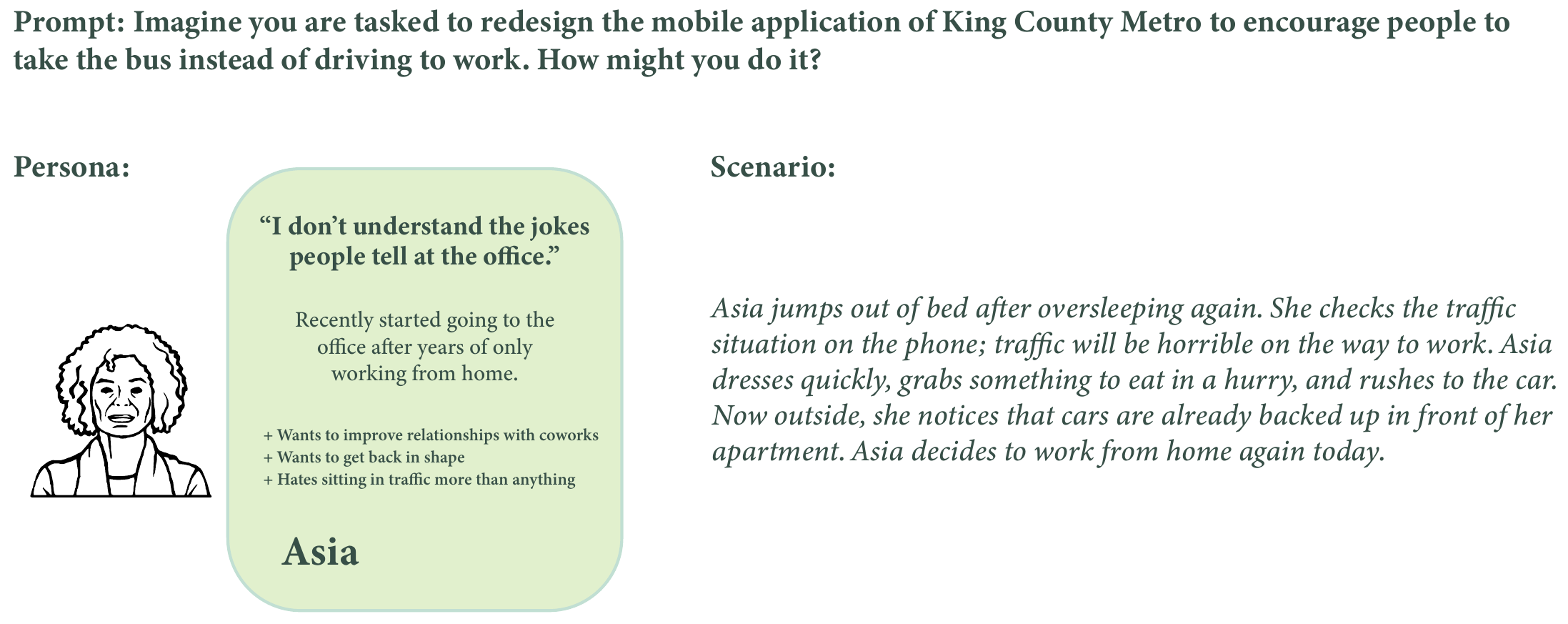}
    
    \caption{An example design sprint}
    \label{fig:designsprint}
\end{figure*}

The study was a one-hour, within-subjects online study. First, we provided participants with an introduction to CPD, explaining its different components, and gave a tutorial on the process of generating a CPD using an example. We then asked participants to complete two 10-minute design sprints. For each design sprint, participants were given a design prompt, a user persona, and a scenario explaining the issues the persona was encountering in that context. \autoref{fig:designsprint} shows an example design sprint. Participants were asked to generate CPDs to ideate possible solutions. We structured our study as design sprints as they are commonly used design methods during early-stage design to explore design ideas~\cite{banfield2015design} and have been used in prior work to support the ideation of theory-driven designs~\cite{colusso2018behavior}. 

For one of the design sprints, participants were given the plugin to help generate CPDs. In the other design sprint, they were not given the plugin. We randomized the order in which a participant would use the plugin, as well as the order of the design sprints to account for any ordering effects. After each design sprint, we asked participants questions about their experiences with using CPD (and the plugin when in the plugin condition). Details of these questions are shared in \autoref{section:study-measures}. After the design sprints, we interviewed participants to further understand their experiences. A detailed version of the protocol can be found in \autoref{appendix:study-protocol}.

\subsection{Recruitment \& Participants}

We recruited participants by posting a screener survey on social media. Of those who expressed interest, we were able to recruit 20 people to participate in our study. The study lasted an hour. All participants except one completed both design sprints. One participant had to leave early due to personal reasons and only completed one of the design sprints (one without the plugin). We provided a compensation of \$50 gift card in compensation for an hour of the participants' time. 13 of our participants are professional UX designers, while others are enrolled in design-related programs. 15 of the participants have at least 2 years of design experience, and all participants had no knowledge of causal pathway diagrams prior to the study. Detailed demographic information of the participants is shared in~\autoref{appendix:demo}.

\subsection{Qualitative Analysis}

Each study session was recorded and transcribed using Zoom. We used thematic analysis on the transcripts. One researcher first used thematic analysis on five transcripts. Then four researchers discussed the excerpts extracted until they decided on a final codebook. The final codebook includes themes such as \textit{CPD facilitates brainstorming in the early stages of design, CPD helps establish a common language between designers}, and \textit{AI recommendations help paraphrase ideas.} The first author then used the codebook to code all 20 transcripts. Throughout the coding process, the researchers mainly focused on how designers used CPD in human-centered design, and how they used the plugin to generate CPDs.

\subsection{Measures \& Statistical Analysis}\label{section:study-measures}

After each design sprint, we asked participants to rate their experience on a Likert scale of 1-7, exploring whether having access to the plug-in influenced their experience generating CPDs. Specifically, we asked them to rate \textit{``how hard/easy it was to design a CPD'',} \textit{``how hard/easy it was to create each component of CPD graphically'',} and \textit{``how hard/easy it was to brainstorm the content of each component of CPD.''} We also asked participants to rate their confidence in the structural correctness (\textit{all components are connected and ordered in the right way}) and content usefulness (\textit{the likelihood of using the generated solution in the next stage of HCD}) of the CPD generated. We analyzed these measures using paired sample t-tests to show how the use of the plugin influenced people's experience using CPD for design.

Apart from the self-reported ratings, we also measured the amount of time participants spent on each design sprint and the number of CPD pathways designers generated for that sprint. We similarly performed paired sample t-tests to analyze the results.
\section{Results}
We focused our analysis on two aspects. First, we identified how the use of CPD facilitated evidence-based HCD. Secondly, we analyzed how our AI-assisted plugin helped UX practitioners more easily work with CPDs, identifying its potential to increase creativity and provide evidence-based support.
\subsection{RQ1: Use of CPD in Human-Centered Design}
\subsubsection{Established an effective and guided design process}

Overall, participants found that utilizing CPD helped them design more effectively for the design sprint challenges. The use of CPD helped direct participants to think more about the constraints involved in the design, as well as the outcome of the design process.

\paragraph{Identify relevant constraints upfront}

Recall that creating CPDs involves mapping out factors of relevance and their relationships (\eg{``what is the distal outcome'', and ``what is the barrier that may occur given this outcome''}). Participants pointed out that when they performed this mapping, they had to think about the relevant constraints upfront. Thinking through all the possible barriers that may prevent strategies from achieving the outcome is a \textit{``realistic and down-to-the-ground strategy''}~(P7), and helps save time in the design process. For instance, P12 articulated that \textit{``instead of thinking about the happy and perfect best case user flow, it is more important to know the constraints.''} P5 also pointed out that \textit{``laying out the barriers, setting up the constraints very very quickly''} immediately highlights the pain point, so that \textit{``addressing these barriers would make the design goals quite tangible when thinking implementation strategies.''} The close attention to the constraints behind a design prompt helps designers quickly break down the task at hand, and helps them focus their effort.

\paragraph{Highlight the desired goal/outcome of the design}

Additionally, participants appreciated the goal-oriented process, rather than free-form or solution-oriented as most existing design processes are (\eg{}, user journey map, whiteboarding, storyboarding, etc.). In many design sessions, practitioners may use whiteboarding to \textit{``simply throw out ideas, then sort out the details, whether an idea is a barrier or a mechanism at a later time.''}~(P7) Or they may use storyboarding, which starts at the beginning, and work from the solution to \textit{``play out how the solution impacts the outcome.''}~(P18) 

In contrast, CPD starts from \textit{``the high-level vision of why focus on this particular problem, the distal outcome,''}~(P6) and \textit{``not just jumping into solutions.''}~(P2) The generation of a CPD emphasizes the answer to the question of \textit{``what is the long-term goal''} and not to \textit{``what are some ways to solve this prompt''}~(P17). Starting the brainstorming process from these questions, the outcome of the design, is actually \textit{``a natural way of thinking, especially when tackling complex problems.''}~(P11) P16 pointed out that one needs to know \textit{``the root cause of the problem''} as well as \textit{``the impact involved in resolving the issue''} before \textit{``identifying the most effective way to address it.''}~(P16) Thus, CPD is effective in facilitating goal-oriented design because it offers \textit{``a direct connection between everything on the users' side to the larger context side, from a visual standpoint.''}~(P6) Furthermore, CPD's emphasis on the outcome helps designers pay attention to\textit{``the realistic and practical goal of the design,''}~(P15) and \textit{``addressing the most fundamental issue in the design context,''}~(P7) instead of thinking about \textit{``designing prettier solutions.''}~(P20)

\subsubsection{Ideation}

When participants were introduced to the concept of CPD, they were intrigued by its potential as a tool for ideation. This is because CPD diagrams can expand into branches, and thus help \textit{``lay out all of the possible solutions''}~(P5) and \textit{``generate a breadth of ideas''}~(P14). Multiple barriers could emerge, each having numerous potential mechanisms and strategies. During the brainstorming sessions, the CPD \textit{``turned into a huge diagram with a bunch of possibilities, moving in so many diverse directions.''}~(P7) Additionally, even after designers have identified a set of barriers and have started working on the design solutions, it is easy to \textit{``backtrack and simply extend another branch''}~(P18) if another barrier pops into their mind.

As designers expand the CPDs in multiple branches, a key step is to think about the causal connections between components. These causal connections kept designers' attention span contained throughout the ideation process. As P19 said, \textit{``creative people are either hyper-focused or are not motivated. But CPD guides you through the brainstorming by asking you to connect from one component to another.''}~(P19) Similarly, P6 pointed out that \textit{``the set of questions highlights the connection between the component I'm working on and the other ones I've created, which kept me on track, dragging me back to the problem space.''} 

However, some participants found that focusing on how the components should be connected to one another at a high level distracted them from brainstorming the individual elements. Specifically, they found that their attention was split between \textit{``ideating additional possibilities of an individual element''} and \textit{``following the step-by-step process to brainstorm the next component.''} (P19) P15 found herself \textit{``moving between the list of questions of the CPD process, jotting down the barrier, but then moving to list out strategies and connecting them to mechanisms.''} Essentially, participants knew that they needed to implement both individual elements and the causal connections eventually. So they tried to multitask but found \textit{``the thoughts scattered across.''} (P2) However, participants also pointed out this issue could be addressed by handling elements and the connections separately. P9 said that she would \textit{``focus on individual elements first,''} then \textit{``come back to the connections in a second design session.''} This process would ensure that one can focus their thoughts on either just component content or just high-level causal relationships.

\subsubsection{Strategic prioritization}

CPDs emphasize the connections between the components, which helps designers play out how a design solution addresses the design prompt, and assist in strategic evaluation. The directional relationships represented by causal connections in CPDs \textit{``served as a guardrail to keep track of why a component is created, how each component is addressing this design scenario.''}~(P14) The \textit{``simple but powerful''}~(P14) causal connections between components make it \textit{``easy to trace back through each pathway and figure out how a solution or a strategy plays out to its mechanisms, resolves the barriers, and achieves its final outcome.''}~(P20)

In addition, since CPD is able to demonstrate how each solution addresses the design problem, participants also used it to compare and prioritize different solutions. Designers shared that CPD is able to present various solutions on multiple branches without becoming too complex. Even as a CPD expands into a large tree with multiple pathways, its main structure retains a \textit{``simple and straightforward''} nature due to the interconnection of components through causal relationships (P11). In addition, the different branches helped designers see \textit{``how the different design solutions are influenced by various barriers,''} which is useful for them to then \textit{``evaluate the likelihood of each barrier, the significance of that barrier, and determine overall which corresponding solution to actually implement and prototype.''}~(P20)

\subsubsection{Facilitating communication}

The use of CPD can also help facilitate communication with stakeholders. P10 pointed out that CPD helps establish a common language among designers in a team setting: \textit{``every designer has their own language, and I've noticed how common it is to have miscommunication issues with others about an idea in brainstorming sessions.''} P20 also said that having a \textit{``clear and straightforward''} process that everyone understands could \textit{``align the way we [designers] see each others' ideas, banding to the same wavelength.''} Furthermore, \textit{``reducing the cost of communication''} means that designers could \textit{``focus more on the ideas themselves and not how to present them.''}~(P5)

Additionally, participants expressed that CPD can bridge the communication gap between designers and the executive team. As discussed before, the process of CPD underscores the goal of the design, which aligns with how product managers and sales consultants think about \textit{``strategizing a product and conducting business-level market research.''}~(P16) P9 also described that \textit{``the distal outcome in CPD''} corresponds exactly to product managers' viewpoint of \textit{``why this problem matters''} and \textit{``what is the long-term business goal of the product.''} Such correspondence means that CPD could easily and effectively communicate designers' ideas to product managers. Furthermore, CPD offers \textit{``the visual clarity for people to easily trace through the logic''} due to its simplistic structure and causal connections (P11). Such a benefit lowers the barrier of communication between designers and product managers.

\subsubsection{Concerns about potential misuse}

Because the CPD process was positively perceived to help with ideation and strategic evaluation, participants expressed concerns that it may be used as \textit{``a tool to oversimplify complex conversations without any proof.''}~(P8) As P20 further explains, \textit{``I'm concerned people would just use this without research, and that could mean putting in random things and coming up with solutions that don’t work.''} Instead, CPD should be used as \textit{``a guided process that helps organize the results of user research.''} (P17) P7 summarized this sentiment with \textit{``A useful tool such as this would only be effective when in the right hands. Simplifying and lowering the barrier to brainstorming is certainly great. But we also need responsible designers who do their due diligence, and do the right user research, before using the CPD to lay out their thoughts.''}~(P7)

\subsection{RQ2: Use of Plugin in Generating CPDs}

\subsubsection{Quantitative results}

\begin{figure*}
    \centering
    \includegraphics[width=\linewidth]{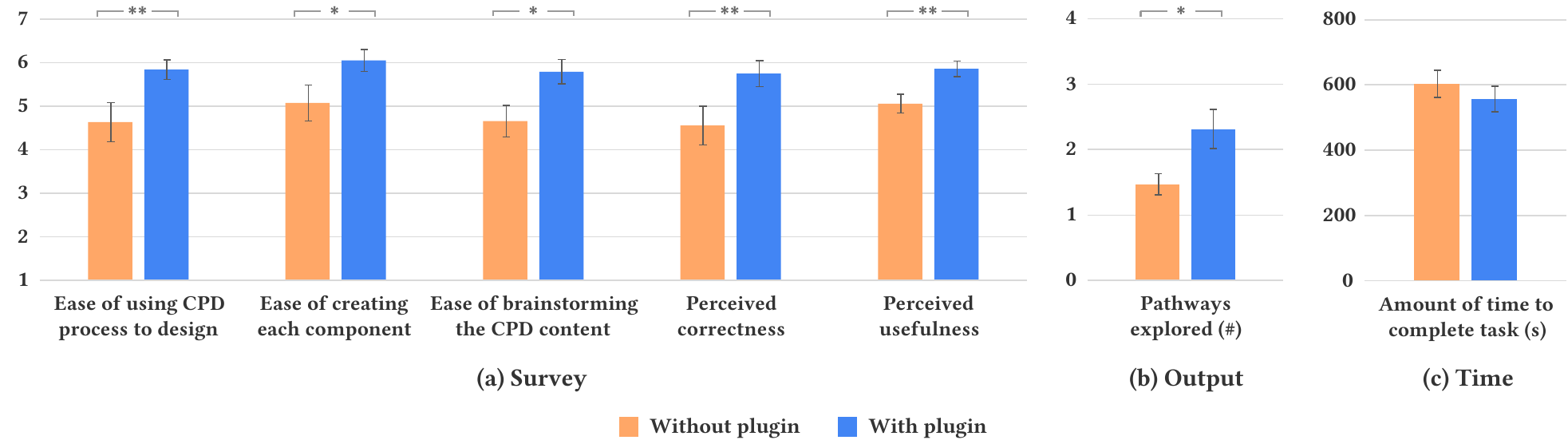}
    \caption{Quantitative results of our user study. Bars indicate standard errors (*: $p<.05$, **: $p<.01$)}
    \label{fig:stats}
    \labelphantom{fig:stats:survey}
    \labelphantom{fig:stats:output}
    \labelphantom{fig:stats:time}
\end{figure*}

Our analysis of the participants' self-reported ratings showed that they had a positive attitude toward the plugin. As shown in \autoref{fig:stats:survey}, their self-reported rating of the ease of creating each component ($t(18)=-2.370$, $p<.05$) and ease of using the CPD process to design ($t(18)=-2.987$, $p<.01$) increased when they had access to the plugin. Additionally, the plugin increased participants' rating of the easiness of brainstorming the content of each component ($t(18)=-2.649$, $p<.05$).

In addition to the self-reported perception towards the use of our plugin, participants' self-rated confidence level of the generated CPD's correctness ($t(17)=-3.409$, $p<.01$) and usefulness ($t(17)=-3.692$, $p<.01$) increased when they had access to the plugin (\autoref{fig:stats:survey}).

\begin{figure*}[t]
\centering
    \begin{subfigure}[b]{\textwidth}
                 \centering
                 \includegraphics[width=0.99\textwidth]{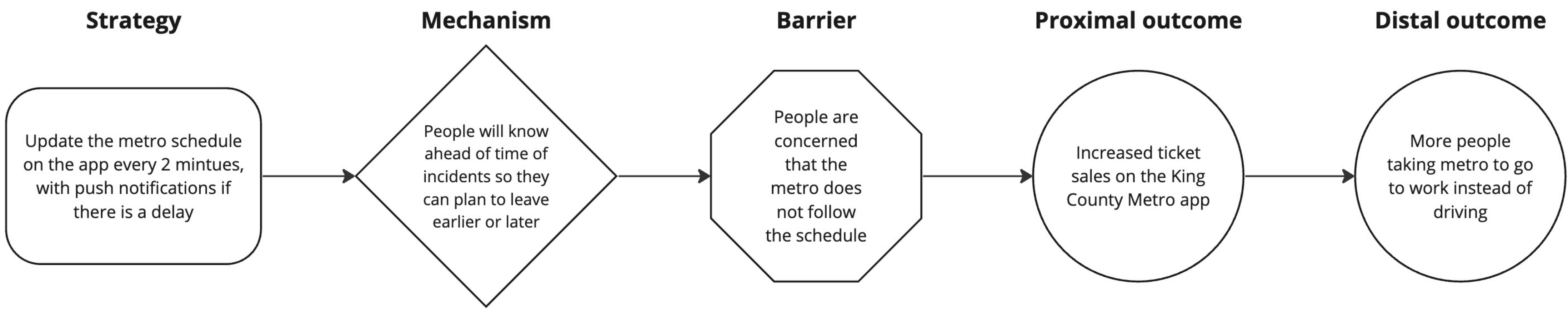}
                 \caption{A CPD generated by P13 under the condition without access to the tool}
                 \label{fig:withoutsample}
    \end{subfigure}\\[1em]
    \begin{subfigure}[b]{\textwidth}
                 \centering
                 \includegraphics[width=0.99\textwidth]{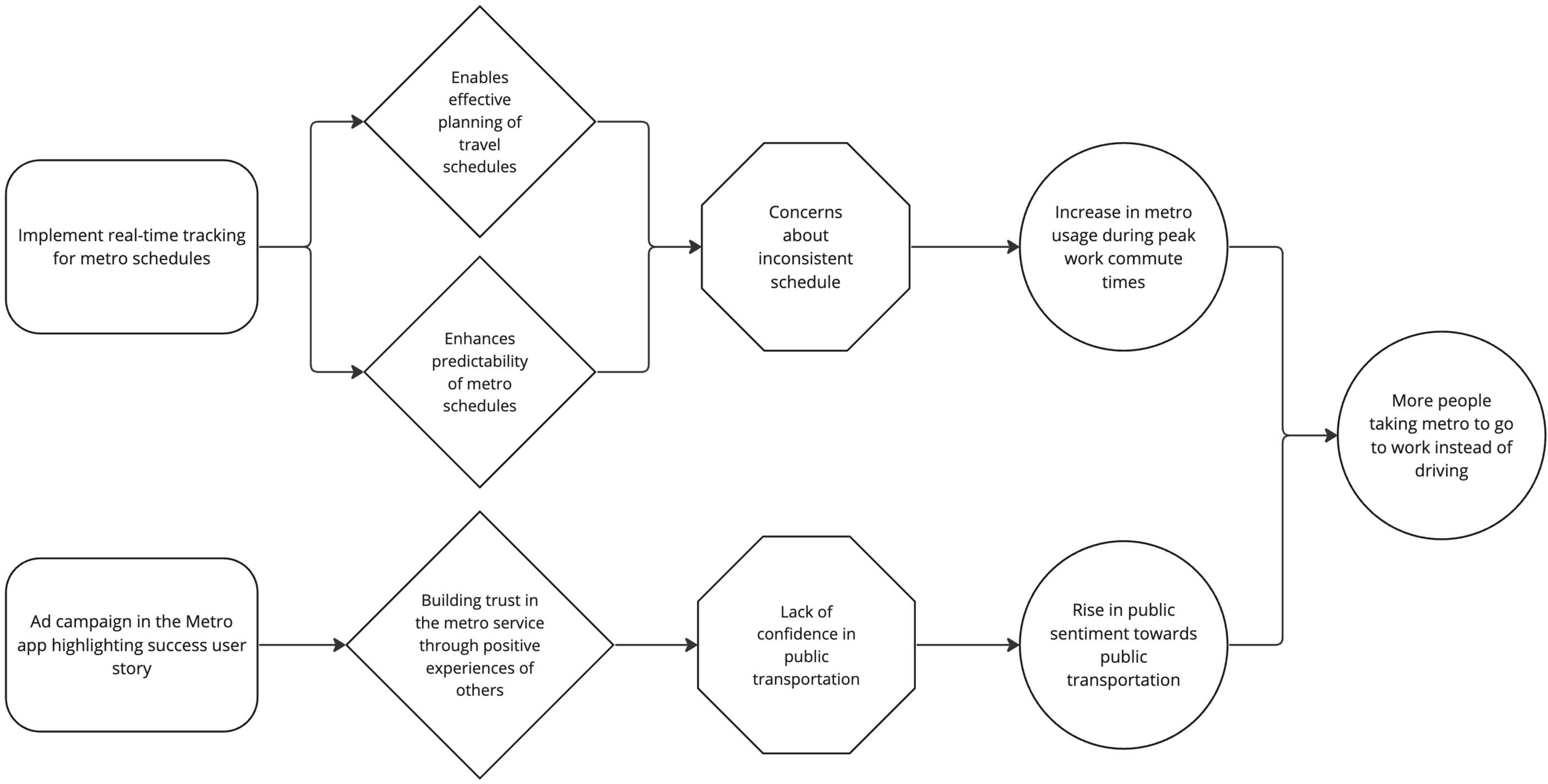}
                 \caption{A CPD generated by P8 under the condition with access to the tool}
                 \label{fig:withsample}
    \end{subfigure}
    \caption{Sample CPDs generated during the design sprints by participants to address the prompt shown in \autoref{fig:designsprint}}
    \label{fig:cpd-participant}
\end{figure*}

When analyzing the CPDs generated, we also found that when using the tool, participants created more pathways in their diagrams ($M_{{without\_plugin}}=1.47$ vs. $M_{{with\_plugin}}=2.32$; $p<.05$), using about the same or less time ($M_{{without\_plugin}}=603(s)$ vs. $M_{{with\_plugin}}=556(s)$; $p > 0.5$). For examples of CPDs generated without and with the tool, see \autoref{fig:withsample} and \autoref{fig:withoutsample}.

Overall, our quantitative analyses demonstrated the plugin helped designers more easily navigate the use of CPD, helped them brainstorm more possibilities, and increased their confidence in the work produced.

\subsubsection{Alleviated cognitive workload}

Generating CPD involves remembering the framework itself, which could distract designers from attacking the problem. The plugin provided the framework details in an easily consumable way and guided users through the process of creating a CPD in an orderly fashion, allowing participants to focus on the design task. By reducing the cognitive workload involved, the plugin not only expedited participants' design process but also increased their confidence in the work they produced.

When participants did the design sprints without any additional help, they expressed that it was particularly difficult to remember the association between each component and its corresponding shapes, such as \textit{``remembering if the shape for the barrier is a diamond or an octagon.''}~(P7) They also struggled with the naming of the components, pointing out that \textit{``looking at the term `mechanism' does not inform much about its meaning and functionality.''}~(P17) Thinking about the CPD framework distracted designers from the ideation process, as they felt like their brain was \textit{``split in half with one focusing on building shapes, the other on the design.''}~(P16) This burden of multitasking was alleviated by using the \textit{Components} feature of the plugin. With an easy drag-and-drop feature where users can generate each component without thinking about its corresponding shape, the tool greatly \textit{``expedites the time''}~(P6) because participants were \textit{``less caught up by the framework itself to actually think about the problem.''}~(P5)

Participants also struggled with how CPD components should be interconnected, including \textit{``what follows after a mechanism,''} or \textit{``whether it is a barrier or strategy that should come up next.''}~(P3) They constantly referred to the generation process before diving into the design details. But the step-by-step process of the \textit{Wizard} feature helped participants \textit{``focus attention on the answer for each component and not worry about the order of them.''}~(P3) P19 also testified that \textit{``bypassing thinking about the ordering made me more empowered to keep searching for a clearer picture of what the design should be.''} Since participants were able to \textit{``think through each component without the disturbance of the shapes or how they should be ordered,''} using the plugin, they felt \textit{``more sure and more confident''}~(P13) of their design. Similarly, P9 said that \textit{``using the Wizard feature, I knew that I followed the right procedure,''} and \textit{``I had more confidence in my design because I spent all my energy on it.''}

\subsubsection{Increased creativity with the support of AI-generated recommendations}

Utilizing LLM, our plugin offered suggestions of component content when prompted in the \textit{Wizard} and the \textit{Brainstorming} features. Details of the prompts used were discussed in \autoref{section:design}. Generally, participants expressed positive attitudes toward using AI-generated content in designing CPD helped them clarify their ideas and sparked innovative thinking during brainstorming. However, they also expressed concerns about blindly using the suggested content.

\paragraph{Increased creativity in brainstorming sessions}
\label{results:creativity}

As discussed in our quantitative findings, the use of the plugin increased the number of pathways designers were able to generate. Participants' comments corroborated this observation and highlighted that some AI-generated content helped increase their creativity. For example, when given the same design sprints (\autoref{fig:cpd-participant}), P13 (without tool) only generated one pathway, while P8 (with tool) not only generated a similar path as P13, but also generated a distinct second path. 
Specifically, P8 commented that the brainstorming feature \textit{``prompted me to think about a new direction as I was thinking about what barriers there are in the problem space.''} And P8 described that he would not have thought about \textit{``the idea of running an ad campaign in a short timeframe if I were just throwing out ideas by myself. But seeing the suggestions by the tool immediately clicked for me. So I also explored this path in the CPD.''} Further, of the path that P13 and P8 generated that were similar, P8 also was able to consider a different mechanism. These differences would have allowed P8 a richer design space (that is still evidence-based) to explore. P9 shared a similar perspective: \textit{``having the AI on the side was like having lots of teammates with really different viewpoints.''} Additionally, AI can be used to expand creativity by using it to weed out the wrong directions: \textit{``In any design, your first 50 ideas are going to be trashed, but you have to get them out to get to idea 51, which is how the AI can help.''}~(P14)

\paragraph{LLM helped with the articulation of ideas}

AI did not always generate new ideas that the participants had not already thought of. When designers' ideas overlapped with AI recommendations, designers found that how AI phrased these ideas was helpful. Participants noted that AI articulates these ideas in \textit{``effective, condensed phrasings,''} or the AI \textit{``uses expressive terminologies (\ie{} route optimization algorithms) to summarize''}~(P3). Reading the AI suggestions helped designers \textit{``build the idea firmly in the mind in a clear and direct way.''}~(P14) AI's articulation and clarification of the concepts helped participants focus more of their energy on the \textit{``meaning and impact of the ideas''}~(P14), rather than \textit{``drilling on how to best describe them.''}~(P20). For example, when communicating the same barrier in their CPD (\autoref{fig:withoutsample} \& \autoref{fig:withsample}), P13 (without tool) had to use a full sentence to describe the barrier: \textit{``people are concerned that the metro does not follow the schedule''} whereas P8 (using the tool) was able to describe it more succinctly \textit{``concerns about inconsistent schedule.''}

\paragraph{Desire for more context-specific/relevant support}

Though participants were generally positive about AI recommendations, they did point out that some of the AI recommendations were too generic and unhelpful, especially when users did not provide sufficient context information in existing components. For example, when P17 inputted (\textit{``improve bus schedule''}) as a generic distal outcome in the \textit{Wizard}, the plugin was not able to create specific and contextualized examples of barriers. Instead, the generated suggestions were  \textit{``oversimplistic and did not seem to fit the problem context.''}~(P17) P14 had a similar issue, where the AI provided out-of-context suggestions when it was given non-descriptive component content to work with. Additionally, P6 said that in his previous encounters with AI recommendations, it is common for the AI to get repetitive, and it had the tendency to constantly resort to \textit{ ``recommending designing an immersive VR experience''} for any given design problem when he did not define it with excruciating detail.

\paragraph{Making recommendations (more) evidence-based}

Another challenge with using the AI recommendations is the lack of information about where the recommendations came from. Participants discussed the importance of determining whether the suggestions actually address the specific problem space. Making that determination requires that the designers \textit{``understand what is the background of this specific concept recommended.''} (P12) For example, in our study, participants were careful about using AI-generated content and only chose to proceed if they had previously studied the concept or knew of its background research. But very often, designers do not know everything about the AI recommendations (\eg{}, \textit{``what is the meaning of the suggestions''}~(P2), \textit{``why is it being recommended''}~(P19), etc.). Participants expressed concerns about what designers may do in these situations. P3 highlighted that \textit{``not knowing the research behind these recommendations would sow doubt in my mind about my design, which is not a good sign.''} P2 also emphasized that there could be significant consequences in product deployment if irresponsible designers just choose to \textit{``click through AI suggestions in Wizard without judging whether it actually fits the problem.''} Participants noted that this could be addressed in the future by providing more context information for each recommendation. For instance, seeing \textit{``external links and research papers to explain the reasoning behind''}~(P18) would help them more easily understand and contextualize the AI recommendations. Plus, as the participants pointed out, seeing \textit{``the background research through external sources''}~(P20) would also validate that the AI is not hallucinating, addressing another common concern that participants shared when they were unfamiliar with the AI-generated content.
\section{Discussion}

In this work, we explored how CPD---originating in implementation science---may be used to support theory-driven design in the domain of HCD. To facilitate its use, we developed a Miro plugin with several CPD guidance features and infused it with AI recommendations powered by an LLM. We then conducted a user study with practitioners, where we found that CPD may be helpful in supporting divergent and convergent work within the early stages of design and demonstrated that our plugin is helpful in reducing overhead cognitive costs and helping guide CPD development. Modern design practice is not purely idiosyncratic as Schön described~\cite{schon1983reflective}, and CPD shows promise in making both divergent and convergent processes more predictable. Below we discuss our interpretations of the study findings in more detail.

First, our work demonstrated that the use of CPD facilitated both divergent and convergent thinking through an evidence-based lens. During the divergent processes of addressing the design sprint (\eg{} ideation), CPD allowed practitioners to easily diverge and expand their innovative thinking. In an ideation context, this can support the create activity as noted in Shneiderman's framework\mbox{~\cite{shneiderman2002creativity}.} With the use of visual elements, designers found it easy to organize their ideas and branch off to explore new factors and new paths. Additionally, backward mapping~\cite{elmore1979backward} enabled designers to start with the desired distal outcome and work backward to explore relevant factors in a step-by-step manner, specifying the causal, moderating, or mediating relationships. CPD helped designers reflect on relevant evidence related to the design problems, reflecting the support of the collecting activity~\cite{shneiderman2002creativity}. Instead of impeding creativity, these ``guardrails'' helped remind designers of the design objective and provided a clear structure for evidence-based thinking through the mapping of outcomes, barriers, mechanisms, and strategies. 



CPD can also support convergent processes in HCD. During our sprints, CPD helped designers with strategic prioritization, converging their ideas to select a solution for the next stage of HCD. After initial brainstorming, the generated CPD may have multiple branches, each suggesting potential solutions. To determine which ones to prototype, designers were able to examine across branches and contrast the effectiveness of each solution. Using the identified goals and constraints, they were able to strategically select the ones they believed to have more influence on the problem context and prioritize the implementation of that solution. In practice, we would expect designers to make these judgements by referring back to relevant user research and optimizing on the paths given their contexts.

Our study also uncovered CPD's potential as a communication tool. Within an ideation context, this is relevant for both the relating and donating activities\mbox{~\cite{shneiderman2002creativity}}, but this type of communication support is also critical throughout HCD. Since each designer often has their own language to describe a design idea and articulate how to approach that idea, it becomes difficult for them to brainstorm together in a team setting. As explained in~\cite{gonen2020tim}, design professionals' drawing practices serve to communicate their ideas instead of illustrating designs accurately. In our study, we observed how CPD supported such a process of visual representation of complex ideas. CPD could help provide an established and easy-to-understand framework. For instance, CPD can guide a team of designers together through the process, starting from the outcome of the design, to barriers, and then to strategy. Throughout the process, CPD's procedure ensures that team members are brainstorming the same component. This suggests a promising future research direction--exploring the use of CPD during team ideation~\cite{wang2017literature, gabriel2016creativity, frich2019mapping} and iteration~\cite{little2010turkit, kim2017mechanical, yu2011cooks}. Additionally, designers highlighted the potential of CPD as a tool to present their ideas to business stakeholders and product managers. They found that the questions a CPD is asking coincide with the questions the executive teams ask them. Future work could further investigate how product managers respond to CPD and how effective CPD may be in facilitating meetings between designers and product managers.


Further, based on our observations, it seems that CPD may also help support designers (and user researchers) to build and communicate their own theories more effectively. In this sense, CPD can support---as been noted in prior work~\cite{colusso2019translational,beck2018theory,redstrom2017making}---the theorizing, which is often done by practitioners in practice. Practice is a type of theorizing~\cite{redstrom2017making}, and different concepts and relationships are uncovered and tested in practice and can be useful if bubble up to research~\cite{gray2014reprioritizing}. This makes CPD more than just a tool to help solve a design problem, but also supports the testing of novel theories from the bottom-up and is in itself a knowledge contribution~\cite{beck2016examining}.


In addition to studying CPD's potential use in HCD, our research also explores the use of a dedicated tool to support CPD usage. We found that our plugin helped designers learn and apply CPD. In particular, the \textit{Components} feature helped them easily generate components without thinking about which shape to use. The step-by-step CPD building \textit{Wizard} feature also guided them through the process without being concerned about how the components should be linked. Participants reported higher perceived ease of use of CPD with our tool. By reducing their cognitive workload, the tool helped designers spend more of their energy on the design itself, and participants reported feeling more confident in the accuracy and usefulness of their CPDs created. One thing to note, however, is that while backward mapping~\cite{elmore1979backward} through CPD provides a logical way of thinking about the design problem, participants had trouble navigating this process and brainstorming individual elements at the same time. Designers found themselves multitasking to brainstorm both individual components and causal connections between components at the same time. This may be addressed by increasing flexibility in our step-by-step wizard tool and allowing users to develop different parts of the causal pathway first. 

Finally, an interesting and important point of potential misuse of CPD in HCD was raised by participants. Part of that perception may stem from general concerns that tools (especially AI-based tools) are making the design too easy---and that may be destroying design~\cite{matthews2023destroy, altavilla2020ai}. There are two things to address. First, we envision CPD to complement design, and not replace. Having good causal pathways alone does not guarantee good designs. Skilled designers are still needed to perform that work. Second, CPD is meant to build on existing user research. As highlighted by existing research~\cite{kun2019creative, wan2023investigating, petridis2023anglekindling}, it is essential to include evidence throughout the design process. The fact that participants expressed concerns that CPD may be misused without proper user research may be both an artifact of our study design (sprints where we gave designers pre-developed personas and scenarios), and an indication that our Miro plugin may be lacking in this respect though we had intended to express motivation for evidence-based design. We need to more effectively prompt and remind designers to incorporate evidence-based insights when building their CPD. Thus, there is also a rich opportunity to explore ways to seamlessly integrate user research outputs into CPD usage so that designers are interfaced with more domain-relevant suggestions. 

\subsection{Supporting Ideation with AI-Assisted Tools}
Overall, our study revealed the potential of adopting LLMs to support ideation, specifically, in brainstorming relevant components in causal pathways. Recent research in HCI has begun to integrate LLMs to help users' ideation processes within various contexts, such as news article writing~\cite{petridis2023anglekindling}, idea machine~\cite{di2022idea}, and creative \& argumentative writing~\cite{lee2022coauthor}. Since LLMs have been trained on vast amounts of information, it is possible that they could simulate human cognition and help offer more relevant suggestions based on the world's collective knowledge ~\cite{schmidt2024simulating}. Our work builds on this body of literature and provides both quantitative and qualitative insights into how LLMs can support ideation.

Quantitatively, our work showed that, with LLM assistance, participants were able to generate and explore more variations of causal pathways. Qualitatively, our participants noted that AI assistance was able to offer suggestions from different viewpoints, similar to working with colleagues. To them, the AI recommendations not only made sense, but were also constructive and diverse, which expanded their perspectives on the problem. This reinforced prior works' discussion of how LLMs could support creativity activities by making diverse recommendations and stimulating innovative ideas\mbox{~\cite{ko2023large, petridis2023anglekindling, wan2023felt, rick2023supermind, druga2023scratch,jeon2021fashionq, angert2023spellburst, mccaffrey2018human, klein2020beyond}}. We also found that LLMs helped rephrase and reframe ideas in a more direct and effective way when AI's recommendations overlap with designers' ideas. In particular, LLMs articulated concepts by drawing on succinct terminologies from their large knowledge database. These terminologies helped clarify designers' thoughts on a conceptual level, which reduced their mental workload on rephrasing and reframing the idea themselves. his finding echos prior research's discussion of how LLMs could help explain users' ideas in a clearer way to facilitate brainstorming\mbox{~\cite{gero2019metaphoria, lawton2023drawing, yuan2022wordcraft, kim2023metaphorian, hou2024c2ideas, zamfirescu2023towards, epstein2022happy, almeda2023prompting, druga2023scratch, di2022idea}}.

Another benefit of using LLMs is that it may be able to help reduce design fixation in the ideation process. Design fixation refers to designers' convergence on one idea over divergently thinking about multiple solutions\mbox{~\cite{youmans2014design}}. Prior works~\cite{jeon2021fashionq, figoli2022ai, mccaffrey2018human, klein2020beyond} have discussed that the use of LLMs increased the number of ideas designers could generate compared to without use, indicating an increase in creativity and decrease in design fixation. As suggested in Section \ref{results:creativity}, we observed a similar phenomenon in our finding. Our tool facilitated the generation of additional ideas for CPD, which resulted in more and richer causal paths for them to explore.

Given these affordances, our AI-assisted plugin was well received by the participants and participants were positive about integrating this type of AI-generated recommendations in their future work. However, participants noted at times the suggested factors were too generic and not relevant enough, suggesting that our tool should be improved to provide more context-specific recommendations. We noted that this primarily occurred when the information provided by participants was too high level to begin with, which resulted in the LLM responding with generic factors that may not be specific to the problem space~\cite{zamfirescu2023johnny, white2023prompt}. Future work should explore how to guide users to generate specific and informative outputs. Specifically, the tool could give users a tutorial on how to best prompt LLMs, or present the users with several examples with the expected level of specificity.


Despite the usefulness of our AI-assisted tool in supporting ideation, our participants did express concerns about generative AI's hallucination problem~\cite{alkaissi2023artificial}. Participants were sometimes hesitant to use the recommendations because they wanted to know why AI was making that suggestion, especially when designers had no background knowledge about the AI-recommended content. This echoes prior research's findings that AI-driven tools should provide suggestions and generated content with support from data to enhance its credibility~\cite{liao2020questioning, mohseni2021multidisciplinary, kun2019creative}. This is a critical issue given that CPD is meant to be evidence-based and theory-driven. Our hope is that in the long run, we can build up a repository of developed and experimentally evaluated causal pathways to power our recommendation to designers. However, because such a repository is yet to exist and even if it exists, may be sparsely populated, we envision there is still value in utilizing LLMs in the process. One possibility is to produce intermediate reasoning steps by leveraging the chain-of-thoughts~\cite{wei2022chain} technique, which may not only serve as an explanation that enables users to examine how reasoning processes unfold but also enhance the actual accuracy of the output~\cite{wei2022chain}. Indeed, recent research showed that strategized prompting can allow generic models to achieve good performance in domain-specific tasks~\cite{bubeck2023sparks}. Another is to have the LLM provide recommendations on a more constrained body of knowledge (\eg{}, ask it to make recommendations from an inputted set of papers or factors). However, even if the information is accurate, it would be important to ensure users' trust in the system. The challenge then becomes more about providing information about where the recommended concepts come from (information provenance), which can both help improve trust and help designers make more informed decisions. 

While our study mainly focused on testing a digital AI-assisted tool in virtual/remote sessions, our tool may also be used to support in-person design sessions. We envision that a facilitator could project our tool on screen and ask participants to discuss and collaborate around it. And whenever needed, participants can leverage the benefits of tangibility in the physical spaces~\cite{yaneva2005scaling, frens2013make, jowers2008supporting, gulay2019integrated}. For instance, working on top of the projected screen showing an initial version of the CPD, participants could use Post-Its to brainstorm additional ideas and lay over the current version. Participants could use Post-Its to lay out simple sketches to increase people's understanding of an idea. They could also use Post-Its to facilitate a voting process for idea selection. Future work can explore how to best support this process in-person session to ensure effective design outputs.
\section{Limitation \& Future Work}\label{section:futurework}

The design sprints we used for our work have both strengths and weaknesses. Although the sprints allowed us to study the use of CPD for HCD in a more controlled setting, these design sprints are stylized design tasks that are often used in the early stages of design. They did not allow us to examine the use of CPD to support design in-situ, and across later stages of design. Additional research is needed to further explore how CPD and our plugin can be integrated into the existing design process and be used to support different phases of design.
\section{Conclusion}

This paper explores the potential synergy between Causal Pathway Diagrams (CPD) and Human-Centered Design (HCD). While CPD traditionally serves as a powerful tool for theory-driven behavioral implementation strategies, our investigation has demonstrated its applicability and benefits in the early phases of HCD. Designers embraced CPD as a means to emphasize goal-oriented design by addressing root causes, particularly for brainstorming and strategic prioritization. To address the conceptual and practical challenges inherent in CPD adoption, we introduced a user-friendly CPD plugin integrated with generative AI capabilities. This tool helped streamline the CPD creation process and encouraged evidence-based thinking and creativity among designers. Our findings shed light on the opportunities and responsibilities associated with integrating AI assistance into creative, evidence-based design practices, offering valuable insights for both the HCD and implementation science communities.

\begin{acks}
Research reported in this publication was supported by the National Cancer Institute of the National Institutes of Health under Award Number P50CA244432. The content is solely the responsibility of the authors and does not necessarily represent the official views of the National Institutes of Health. We also thank Bryan Weiner, Cara Lewis, Aaron Lyon, and other members of the OPTICC Center.
\end{acks}
\bibliographystyle{ACM-Reference-Format}
\bibliography{10-bibliography}

\appendix

\section{Demographics}
\label{appendix:demo}
\autoref{table:demographics} shows the demographic information of participants in the study:
\begin{table}[h!]
\caption [Short Heading]{\protect Demographics information of participants. * indicates the participant only completed one of the design sprints}
\label{table:demographics}
\begin{tabular}{lllll}
\toprule
\textbf{PID} &
  \textbf{\leftcell{Age\\range}} &
  \textbf{Gender} &
  \textbf{Occupation} &
  \textbf{\leftcell{Experience\\with design}} \\ \midrule
P1  & 30-39        & \leftcell{Prefer\\not to say}  & Graduate student                  & 2-5 years                       \\
P2  & 24-29        & M  & Product designer                & 2-5 years                      \\
P3  & 18-23        & F  & Graduate student               & 1 year                 \\
P4  & 18-23        & \leftcell{Gender-\\non\\conforming}  & Graduate student                 & 1-2 years                 \\
P5  & 24-29        & M  & UX designer             & > 5 years                  \\
P6  & 24-29        & M  & UX designer       & 2-5 years                      \\
P7  & 30-39        & M  & UX designer   & > 5 years         \\
P8  & 30-39        & M  & Graduate student                 & 2-5 years                       \\
P9 &
  24-29 &
  F &
  \leftcell{Pharmaceutical\\scientist}
  &
 2-5 years   \\
P10 & 18-23        & M  & UX designer      & 2-5 years     \\
P11 & 18-23        & F  & UX designer       & 1-2 years    \\
P12 & 24-29           & F & Graduate student          & 2-5 years                      \\
P13  & 24-29           & F & Graduate student           & 2-5 years        \\
P14 & 24-29        & M  & UX designer               & > 5 years    \\
P15* &
  N/A &
  F &
  UX designer &
  2-5 years \\
P16  & 24-29        & F  & Product designer                  & 1-2 years                     \\
P17 &
  24-29 &
  M &
  UX designer
  &
  2-5 years   \\
P18  & 30-39        & F  & UX designer & 2-5 years                      \\
P19  & 24-29        & M  & Product Designer                 & 2-5 years                    \\
P20  & 24-29        & F  & UX designer                  & 1-2 years     \\
\bottomrule
\end{tabular}
\end{table}

\section{User Study Protocol}

\label{appendix:study-protocol}
\subsection{Phase 1}
\begin{itemize}
    \item Study Introduction 
    \item Introduce CPD with an example
    \item Background questions
    \begin{itemize}
        \item What are your prior experiences with Miro? What do you usually use it for?
        \item How much do you know about CPD?
    \end{itemize}
\end{itemize}

\subsection{Phase 2}
Invite participant to Miro board without plugin
\begin{itemize}
    \item Provide 10 minute design challenge
    \item Ask participant to generate a CPD depicting the challenge
    \item Ask participants to complete the post-task short survey
    \item Answer the following questions on a scale of 1-7
    \begin{itemize}
        \item How hard/easy was the process to design a CPD?
        \item How hard/easy was it to create each component (\eg{}, Strategy, Mechanism, etc.)?
        \item How hard/easy was it to brainstorm the content of each component to create the CPD (\eg{}, Strategy, Mechanism, etc.)?
        \item How confident are you about the correctness of the CPD designed?
        \item How confident are you about the usefulness of the CPD designed?
    \end{itemize}
\end{itemize}

\subsection{Phase 3}
Invite participants to Miro board with plugin (install the plugin)
\begin{itemize}
    \item Briefly go over plug-in features
    \item Provide 10 minute design challenge
    \item Ask participant to generate a CPD depicting the challenge
    \item Ask participants to complete the post-task short survey, which is the same one as used in Phase 2
\end{itemize}

\textbf{Note: We would randomize the order of phase 2 and phase 3 for each participant. 
}

\subsection{Phase 4. Follow-up interview }
\begin{itemize}
    \item Was it difficult applying CPD into this design problem?
    \item Was it helpful using CPDs to tackle these design challenges?
    \item Would you use CPDs in your future design work? Why/Why not?
    \item Did the Miro Plug-in help or impede your ability to create CPDs? In what ways?
    \item What did you think about each feature? 
    \item Was it difficult to use? Did it make a difference when you were creating the CPD?
    \item Do you envision any changes to the feature?
    \item Did the Miro Plug-in help or impede your ability to design? In what ways?
    \item What did you think about each feature?
    \item Did it make a difference when you were designing?
    \item Are there any additional features you think might be useful that might help in creating the CPD?
\end{itemize}

\section{Key Screens of Our Plugin}
\label{appendix:keyscreens}

\begin{figure*}[h]
    \centering
    \includegraphics[width=.87\linewidth]{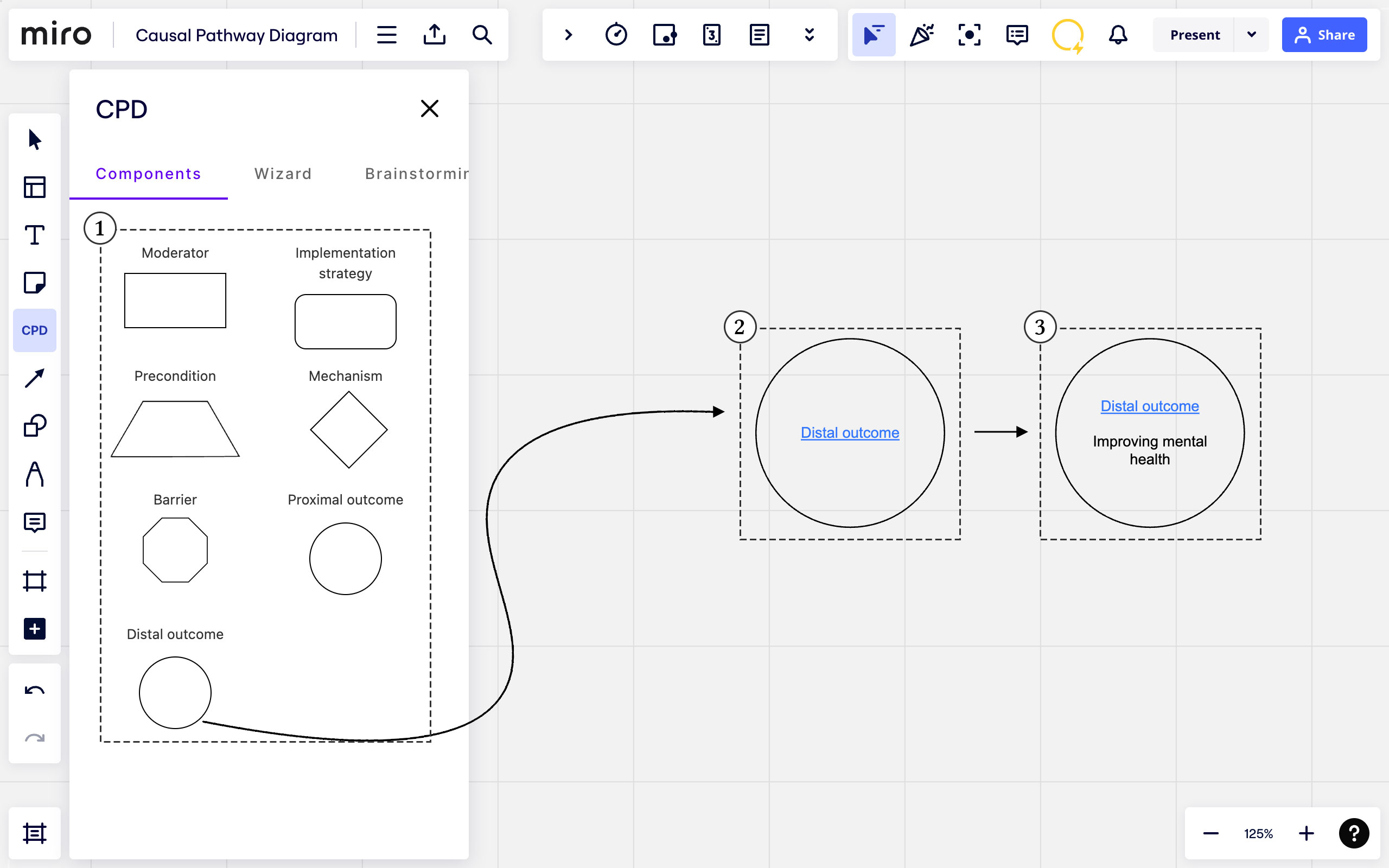}
    \caption{Interaction flow of the ‘Component’ feature. From the (1) plugin panel, users can (2) drag \& drop each component to the board and (3) type the content inside it.}
    \label{fig:enter-label}
    \vspace{.3cm}
    \centering
    \includegraphics[width=.87\linewidth]{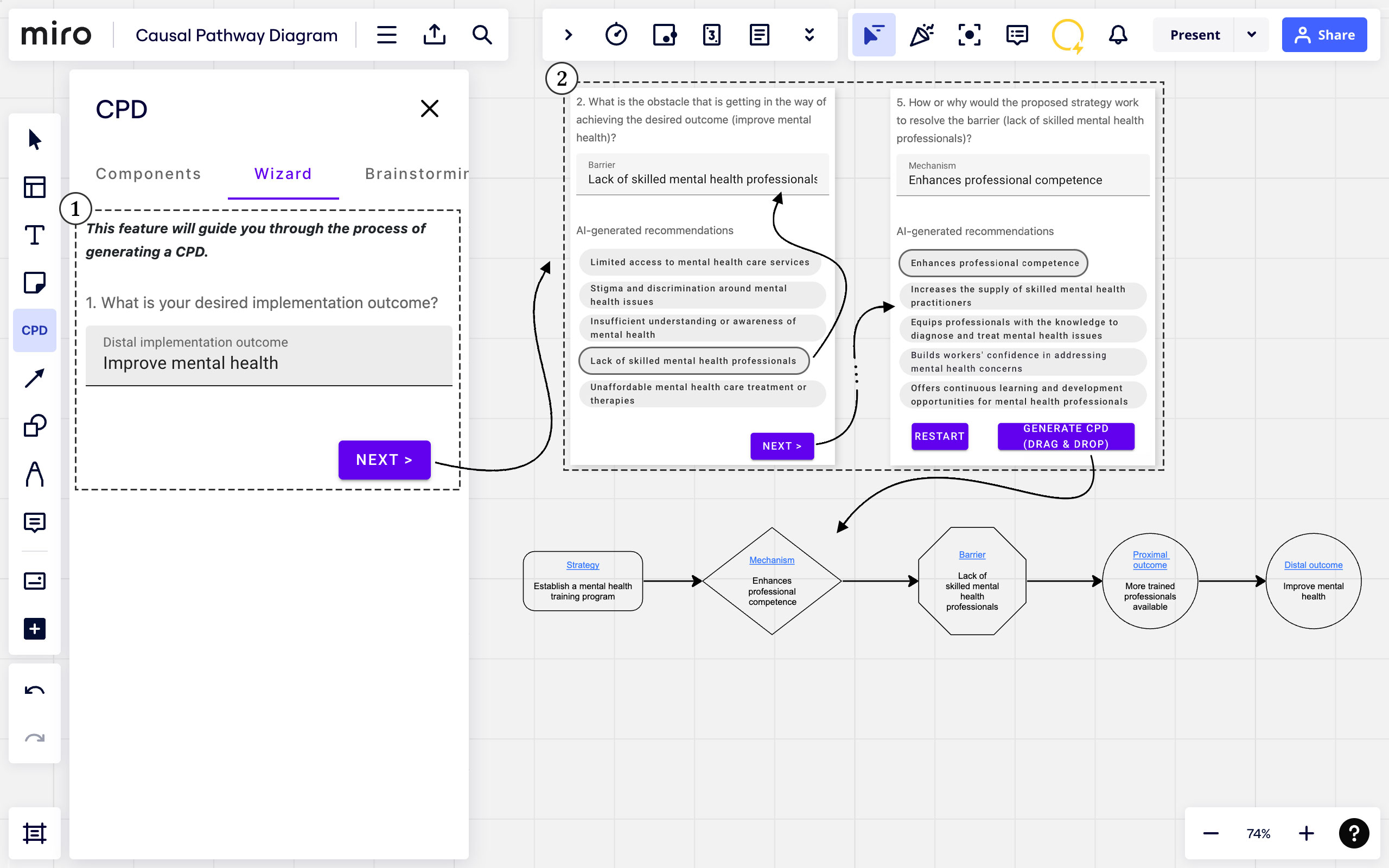}
    \caption{Interaction flow of the ‘Wizard’ feature. Once the user first (1) provides their desired implementation outcome, the system (2) guides them through the other components, along with providing AI-generated recommendations to help them generate CPDs. Once the user types/selects contents for every component, the feature prompts the user to drag \& drop to which they want to place their CPD.}
    \label{fig:enter-label}
\end{figure*}

\begin{figure*}[h!]
    \centering
    \includegraphics[width=.87\linewidth]{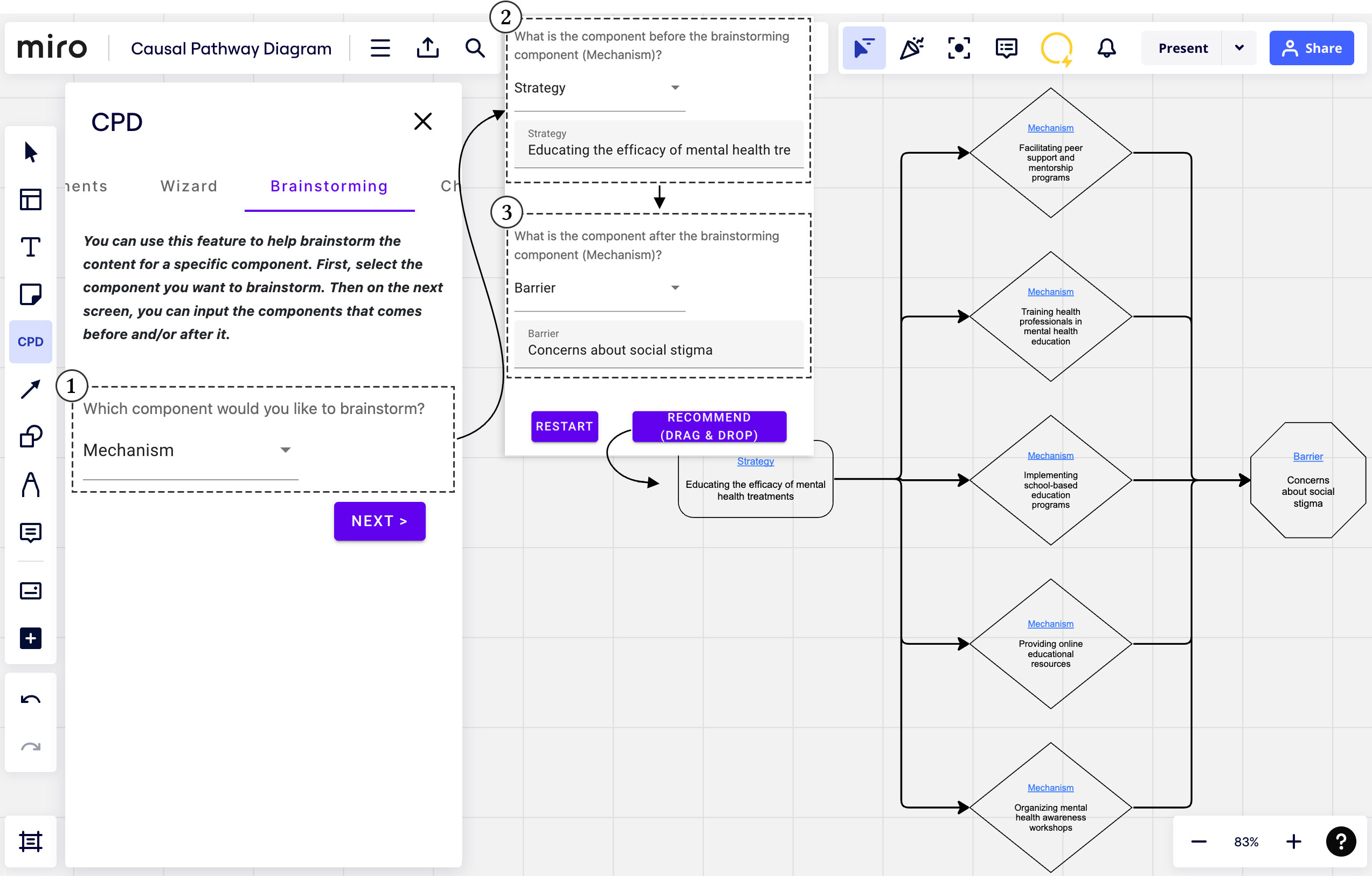}
    \caption{Interaction flow of the ‘Brainstorming’ feature. The plugin first asks the user to provide (1) the component they would like to brainstorm and (2) the preceding / (3) the following components of it. Then, the feature prompts the user to drag \& drop to the board, which then generates five recommendations for the component.}
    \label{fig:enter-label}
    \vspace{.5cm}
    \centering
    \includegraphics[width=.87\linewidth]{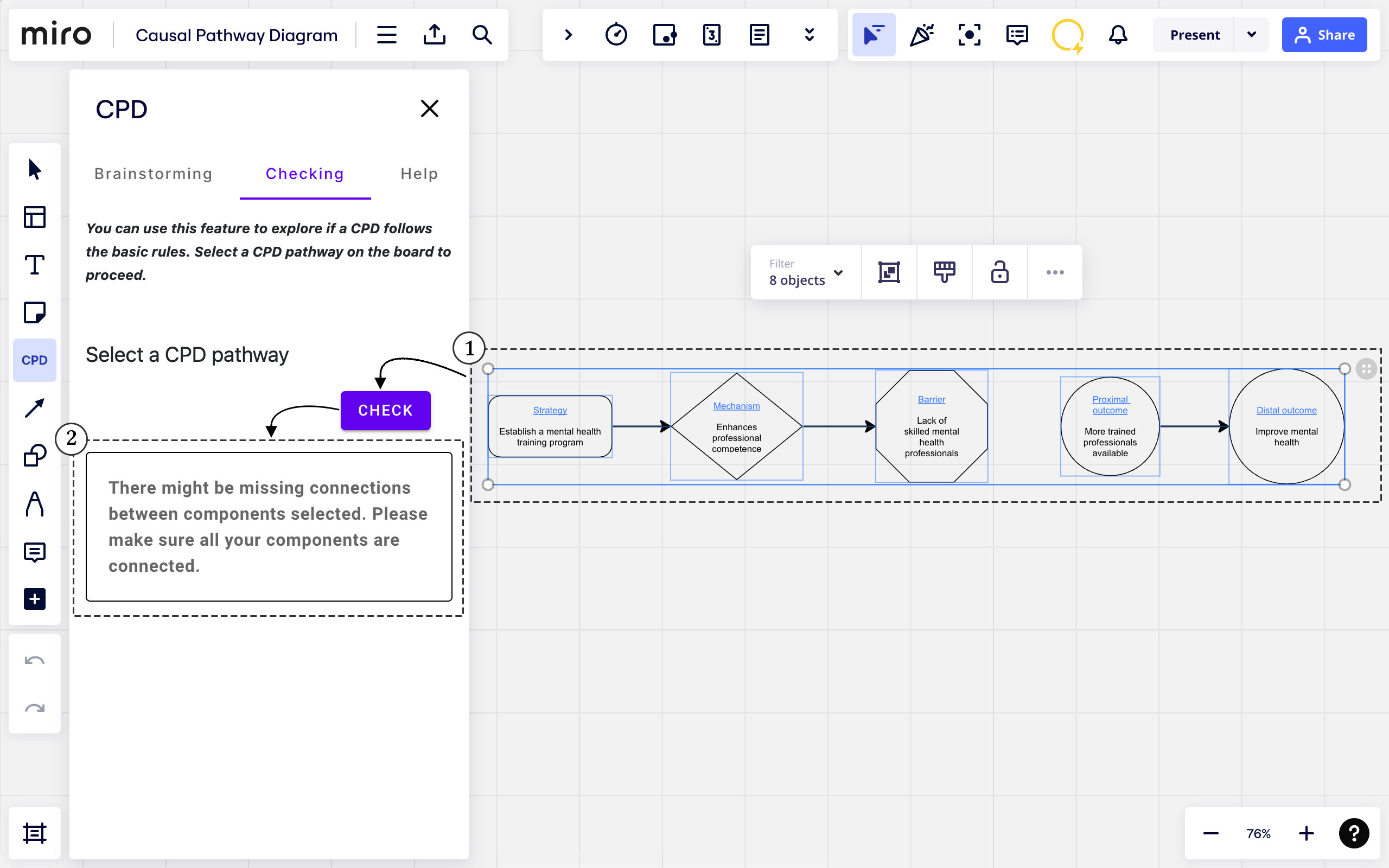}
    \caption{Interaction flow of the ‘Checking’ feature. (1) Once the user selects the CPD that they want to check and clicks the ‘Check’ button on the panel, (2) the plugin detects and shows any potential issue (if any)}
    \label{fig:enter-label}
\end{figure*}

\begin{figure*}[h!]
    \centering
    \includegraphics[width=.87\linewidth]{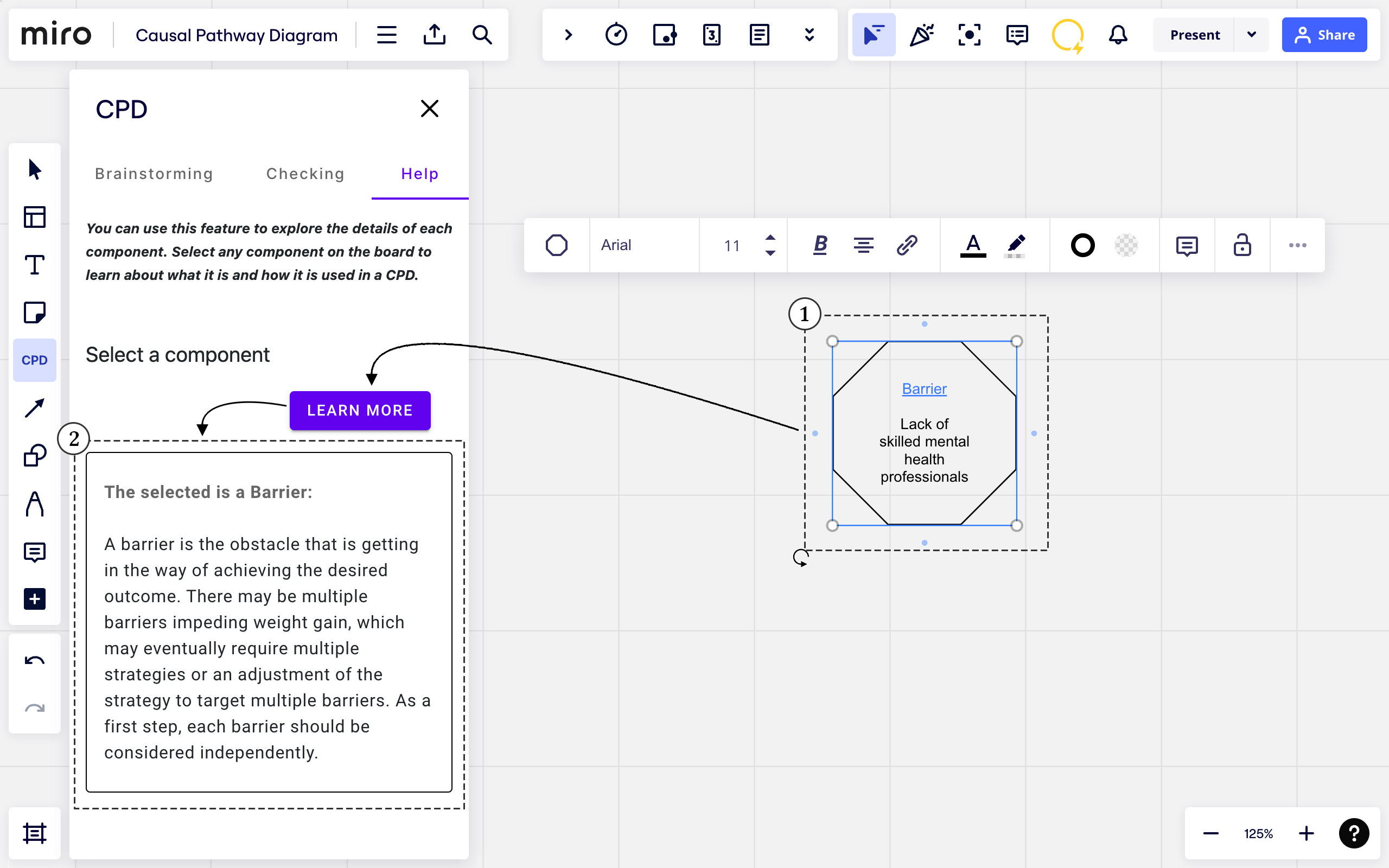}
    \caption{Interaction flow of the ‘Help / Glossary’ feature. (1) Once the user selects the component that they want to learn more about and clicks the ‘Learn more’ button on the panel, (2) the plugin shows the details of the selected component.}
    \label{fig:enter-label}
\end{figure*}

\end{document}